\documentclass[journal]{IEEEtran}
\usepackage{cite}

\ifCLASSINFOpdf
  \usepackage[pdftex]{graphicx}
\else
  \usepackage[dvips]{graphicx}
\fi
\usepackage[cmex10]{amsmath}
\usepackage[caption=false,font=footnotesize]{subfig}
\begin{document}
%
\title{Design and Testing of Superconducting Microwave Passive Components for Quantum Information Processing}
%
%
%


\author{H.~S.~Ku, F.~Mallet, L.~R.~Vale, K.~D.~Irwin, S.~E.~Russek, G.~C.~Hilton, and K.~W.~Lehnert
\thanks{Manuscript received 3 August 2010. This work is supported by the DARPA/MTO QuEST program.}
\thanks{H.~S. Ku, F. Mallet, and K. W. Lehnert are with the JILA, a joint institute between the University of Colorado and the National Institute of Standards and Technology, and with the Department of Physics, the University of Colorado, Boulder, CO 80309-0440 USA (phone: 303-492-7387; fax: 303-492-5235; email: hsiang-sheng.ku@colorado.edu, francois.mallet@colorado.edu, konrad.lehnert@jila.colorado.edu)}%
\thanks{L.~R.~Vale, K.~D.~Irwin, S.~E.~Russek, and G.~C.~Hilton are with the National Institute of Standards and Technology, Boulder, CO 80305-3337 USA (email: leila.vale@nist.gov, kent.irwin@nist.gov, stephen.russek@nist.gov, gene.hilton@nist.gov)}}

%
%

\markboth{2EY-05}%
{2EY-05}
%



\maketitle

\begin{abstract}
We report on the design, fabrication and testing of two superconducting passive microwave components, a quadrature hybrid and a 20~dB directional coupler. These components are designed to be integrated with superconducting qubits or Josephson parametric amplifiers and used in quantum information processing applications. For the coupler, we measure return loss and isolation $\mathbf{> 20}$~dB, and insertion loss $\mathbf{< 0.3}$~dB in a 2~GHz band around 6~GHz. For the hybrid performance, we measure isolation $\mathbf{> 20}$~dB and insertion loss $\mathbf{< 0.3}$~dB in a 10\% band around 6.5~GHz. These values are within the design specifications of our application; however, we find a 7\% difference between the designed and measured center frequency for the hybrid.
\end{abstract}


%
\IEEEpeerreviewmaketitle

\section{Introduction}
%
%
%
%
\IEEEPARstart{M}{icrowave} signals and superconducting quantum bits (qubits) provide a promising architecture for realizing quantum computing and related quantum information processing tasks. To fully exploit the power of microwave circuits, it is important to be able to combine and split propagating modes of microwave fields. These tasks are accomplished using passive microwave components, particularly directional couplers and hybrids, which act as beam splitters for microwave photons.  These linear optics elements provide the ability to manipulate quantum states of light. For example, a directional coupler can be used as a tool to shift a quantum state of light in phase space and a two-mode squeezed vacuum state can be generated at the outputs of a 50:50 beam splitter with two single-mode squeezed states as inputs\cite{Braunstein:RevModPhys:2005}. Although passive microwave components are routinely combined into monolithic microwave integrated circuits\cite{Wen:1970,Ho:1993}, they are only now beginning to appear in integrated quantum information processing circuits. The requirements of a quantum processor yield different optimization of these circuits\cite{Bozyigit:arxiv:2010}.

In this paper, we describe the design and testing of two microwave components, a 20~dB directional coupler and a quadrature hybrid, which are suitable for integration with superconducting qubits and related devices. These components are built to meet the requirements of a particular continuous variables quantum information processing scheme, which exploits the squeezed states prepared by Josephson Parametric Amplifiers (JPAs)\cite{Manuel:NaturePhys:2008}. As such, they must have low loss and high isolation, but only in a 10\% band around a specific frequency of 7~GHz. While these design constraints are specific to a particular scheme, in general quantum circuits are relatively narrowband, but exceptionally intolerant to loss. Superconducting qubit circuits and JPAs are necessarily built from superconducting metals, operated at low temperatures, and often use a CPW architecture.

\section{Design and fabrication of components and calibration standards}

\begin{figure*}[!t]
\centerline{\subfloat{\includegraphics{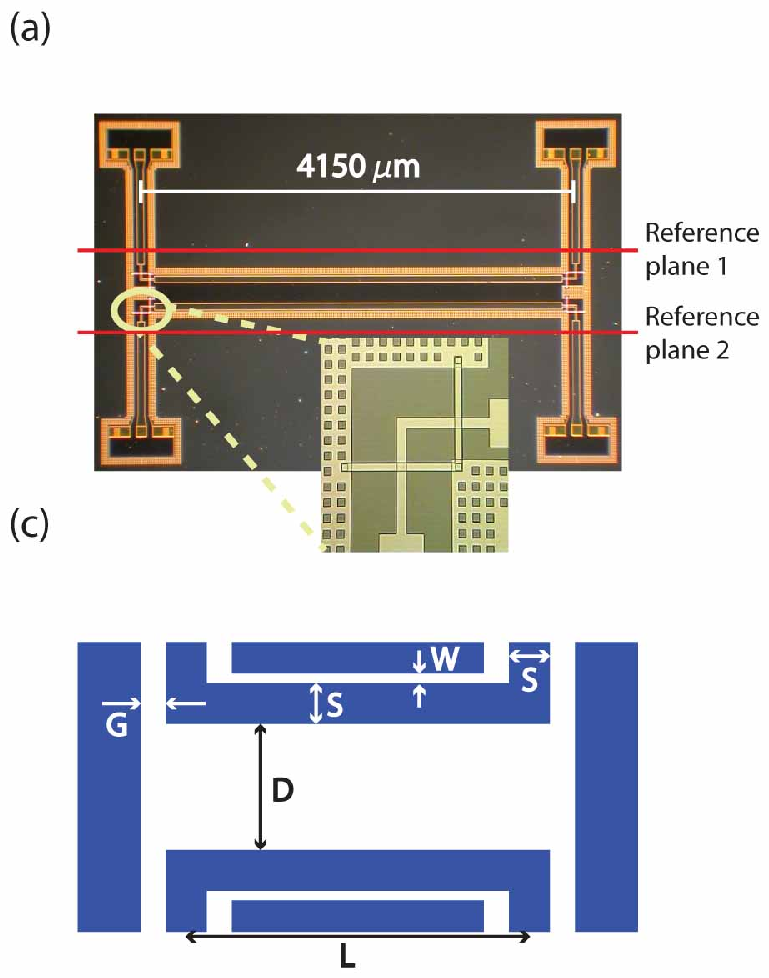}%
\label{fig_20dBdirectionalcoupler}}
\hfil
\subfloat{\includegraphics{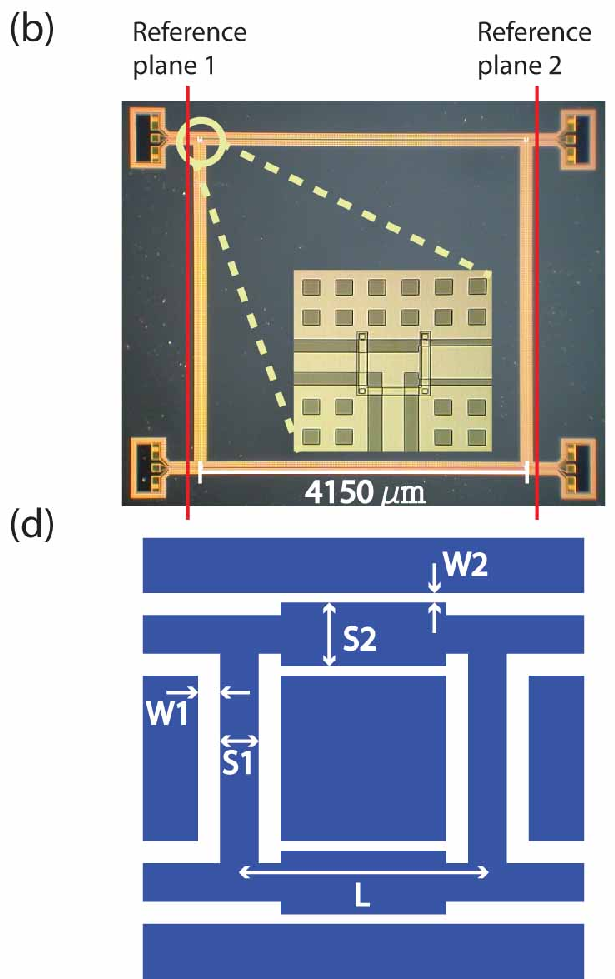}%
\label{fig_quadraturehybrid}}}
\caption{(a) A photograph of the 20~dB directional coupler. The inset shows bridges and compensating structures at the corners as well as ground plane meshes to suppress the motion of trapped magnetic flux. In the inset the lighter material is niobium and the darker material is the insulating substrate. (b) A photograph of the quadrature hybrid. The inset shows similar bridges and ground plane meshes. (c) A schematic of the directional coupler layout. The dark areas represent superconducting metal and the white areas indicate the regions where the metal is removed exposing insulating substrate. These diagrams are not to scale, the indicated dimensions are: $D = 200\ \mathrm{\mu m}$, $S = 65\ \mathrm{\mu m}$, $W = 15\ \mathrm{\mu m}$, $G = 40\ \mathrm{\mu m}$, $L = 4150\ \mathrm{\mu m}$. (d) The layout of the quadrature hybrid with indicated dimensions: $S1 = 14\ \mathrm{\mu m}$, $W1 = 8\ \mathrm{\mu m}$, $S2 = 23\ \mathrm{\mu m}$, $W2 = 3.5\ \mathrm{\mu m}$, $L = 4150\ \mathrm{\mu m}$.}
\label{fig_passivecomponents}
\end{figure*}

In order to be integrated with qubits and JPAs, we design directional couplers and quadrature hybrids using a coplanar waveguide (CPW) architecture from superconducting metals. The CPW architecture is well suited to the the fabrication techniques used to create superconducting integrated circuits. Unfortunately, CPW lines, which support so called quasi-TEM modes, are not ideal for passive components. For this reason, the passive components will have the desired performance over a relatively narrow band, and therefore, must be designed carefully.

We begin by considering the directional coupler, which is realized simply by coupling a pair of CPW transmission lines \cite{Pozar:2005:DirCou,Simons:2001:DirCou}. Fig.~\ref{fig_passivecomponents}(a)~and~(c) show the layout of the directional coupler. The coupler is designed to be matched to 50~$\Omega$ and to have $-20$~dB coupling at 7~GHz. To achieve good isolation, the dimensions of the couplers are chosen such that the difference between the even-mode and odd-mode phase velocities is smaller than 1\%. Bridges in a second metal layer, designed to short the two ground planes of the CPW, are placed at CPW 90 degree bends to suppress the parasitic slot line mode. Furthermore, the narrowed CPW center conductors are used to compensate the parasitic capacitance added by the bridges and the bend reactance\cite{Weller:1998}.

Because it is extremely difficult to achieve $-3$~dB coupling by using coupled-line CPWs, the branch-line coupler structure \cite{Pozar:2005:QuaHyb,Simons:2001:QuaHyb} is used to implement the quadrature hybrids. Fig.~\ref{fig_passivecomponents}(b)~and~(d) show the design of the quadrature hybrids. The initial dimensions of the hybrids are chosen such that the impedances of the through lines and the branch lines are 35~$\Omega$ and 50~$\Omega$, respectively. Both the branch and through lines are designed to be a quarter wavelength long at 7~GHz. As in the design of directional couplers, bridges are placed at T-junctions  to suppress the slotline modes of CPW lines.

\begin{figure}[!t]
\centering
\includegraphics[width=2.8in]{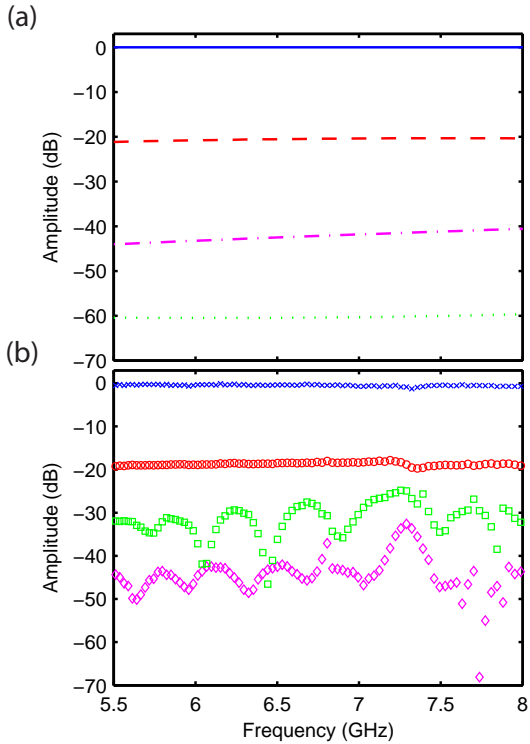}
\caption{(a) The magnitude of the simulated S-parameters of the 20~dB directional coupler showing the fraction of the input power that exits the: through port (blue solid line), coupled port (red dashed line), isolated port (magenta dash-dot line), input port (green dotted line). (b) The magnitude of the measured S-parameters of the 20~dB directional coupler: through transmitted (blue cross), coupled (red circle), isolated (magenta diamond), reflected (green square).}
\label{fig_DirCouAm}
\end{figure}

\begin{figure}[!t]
\centering
\includegraphics[width=2.8in]{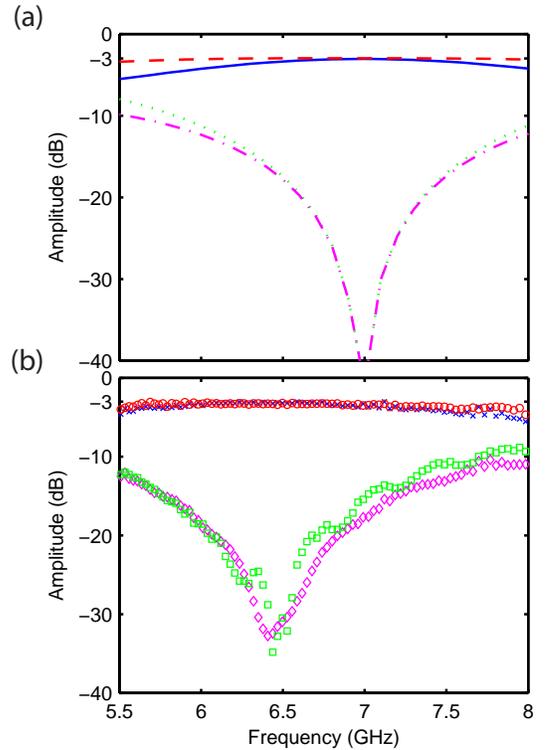}
\caption{(a) The magnitude of the simulated S-parameters of the quadrature hybrid showing the fraction of the input power that exits the: through port (blue solid line), coupled port (red dashed line), isolated port (magenta dash-dot line), input port (green dotted line). (b) The magnitude of the measured S-parameters of the quadrature hybrid: through transmitted (blue cross), coupled (red circle), isolated (magenta diamond), reflected (green square).}
\label{fig_QuaHybAm}
\end{figure}

We optimize the designs of the hybrid and coupler by simulating their performance using AWR's Microwave Office software\cite{nonendorsment}. We then make fine adjustments to the layout dimensions to compensate the effect of the 90 degree bends in the directional couplers and the T-junctions in the quadrature hybrids. We repeat until the simulations indicate that the desired performance has been achieved. For the directional coupler, we can simulate and optimize the entire structure. In contrast, the quadrature hybrid is too complex for the entire structure to be simulated. To overcome this problem, the hybrid is divided into eight parts: 4 T-junction elements, 2 through line elements, and 2 branch line elements. After simulating each element separately to determine its S-parameters, the elements are combined and simulated as a microwave network. Assigning numbers to the labeled ports in Fig.~\ref{fig_passivecomponents}(c)~and~(d) as: input $\to1$, through $\to2$, coupled $\to3$, and isolated $\to4$, we define: return loss $= -20 \log \lvert S_{11} \rvert$, through transmission $= 20 \log \lvert S_{21} \rvert$, coupling $= 20 \log \lvert S_{31} \rvert$, isolation $= -20 \log \lvert S_{41} \rvert$, and insertion loss $= -20 \log ( \lvert S_{21} \rvert + \lvert S_{31} \rvert )$. For the directional coupler, the designed coupling is $-20\pm1$~dB with both return loss and isolation better than 30~dB from 6 to 8~GHz (Fig.~\ref{fig_DirCouAm}(a)). For the hybrid, the designed transmission through and coupling are both $-3\pm0.3$~dB in a 700 MHz band around 7 GHz. The return loss and isolation both are better than 20~dB in the same band. (Fig.~\ref{fig_QuaHybAm}(a)).

A general problem when making microwave network measurements is how to extract intrinsic performance of the device from the frequency dependent behavior of the instruments and cables used to test the devices. The procedure for deembedding the behavior of the measured device from its testing apparatus is called ``calibration.'' The common Short-Open-Load-Through (SOLT) calibration method for coaxial transmission lines is difficult to apply in CPW structures because it requires three lumped elements each with accurate impedance. Instead, the Thru-Reflect-Line (TRL) technique can be used to calibrate devices fabricated in CPW architecture\cite{Pozar:2005:TRLcal,Engen:1979}.

\begin{figure}[!t]
\centering
\includegraphics[width=2.8in]{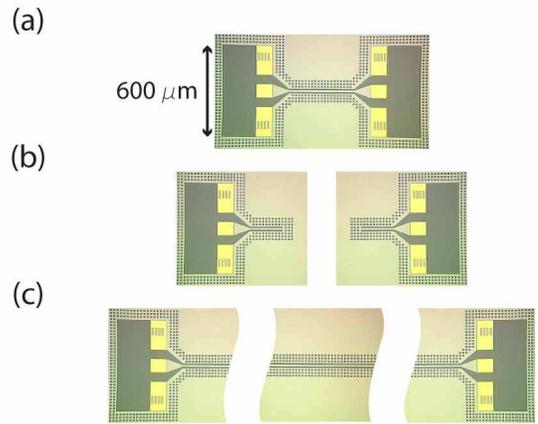}
\caption{The custom TRL calibration standards for the quadrature hybrid. (a) The Thru standard (b) The Reflect standard: two identical CPW short circuits. (c) The Line standard. A similar but different set of TRL standards were fabricated for the coupler (not shown).}
\label{fig_TRLcalibrationkits}
\end{figure}

In order to implement the TRL calibration, we cofabricate sets of TRL calibration standards with the passive components on the same chip. The calibration standards comprise three elements (Fig.~\ref{fig_TRLcalibrationkits}): a section of transmission line providing the ``Thru'' element, a short circuit providing the ``Reflect'' element, and a longer section of transmission line, which is a quarter wavelength longer than the ``Thru,'' providing the ``Line'' element. After measuring the uncalibrated performance of a passive component and the TRL elements, and then applying the calibration algorithm, we can extract the microwave performance of the structure that lies between two so-called reference planes, as shown in Fig.~\ref{fig_passivecomponents}(a)~and~(b). (In TRL calibration, the reference plane is defined by the symmetry plane of the Thru.) Because our custom TRL standards are fabricated with the same probe-pads and tapered CPWs as the passive components, the reference planes on the passive components do not enclose these structures, allowing us extract the true microwave performance of the passive components.

Because the passive components are intended to be integrated with JPAs, our test devices are fabricated with the same niobium process\cite{Sauvageau:1995} as was recently used to fabricate JPAs\cite{Manuel:NaturePhys:2008}. The circuit pattern layer is a 200 nm thick niobium film deposited onto a high resistivity silicon wafer (resistivity $> 17$~k$\Omega\,$cm) with 20~nm of $\mathrm{SiO_2}$ thermally grown on its surface. A 350 nm thick $\mathrm{SiO_2}$ layer is then deposited and etched to form the insulation layer between the circuit layer and a wiring layer of 300~nm thick niobium. Bridges are patterned into the wiring layer, where niobium vias through the insulation connect the circuit layer and wiring layer. For reliable contact between the microwave probes and test devices, a gold layer is deposited on the niobium circuit layer in the region that forms the probe pads. Alignment marks are patterned into that gold layer so that probes can be positioned on the pads reproducibly. In addition to the test devices and the TRL calibration standards, a high $Q$ half-wavelength CPW resonator is also fabricated on the chip for the purpose of diagnosing conductor and dielectric losses. Including all of these elements, the size of the test chip  is 1.5~cm~$\times$~1.9~cm.

\section{Testing and calibration}
The microwave performance of the passive components is measured with a cryogenic microwave probe station system (Desert Cryogenics TT-Prober System) and a vector network analyzer (Agilent N5230A PNA-L)\cite{nonendorsment}. In order to calibrate the passive components with an accuracy of 0.3~dB, a cryogenic microwave probe station is essential. Microwave probes contacting the probe pads provide a high-quality, reproducible connection between the network analyzer cables and the test devices. The quality and reproducibility of this contact is critical because the calibration procedure assumes that the contact made to the TRL standards is identical to the contact made to the passive components. Of course, the probe station must be cryogenic (4~K) to study the niobium passive components in their superconducting state.  The station has four arms connected to four microwave cryogenic probes (Picoprobe P-10-5325-4 (straight probe) and Picoprobe P-10-5325-B ($90^{0}$-probe)\cite{nonendorsment}, so called GSG probes, which naturally launch CPW modes). Because the four probe arms are $90^{0}$ apart from one another, two $90^{0}$ probes and two straight probes are used to match the distribution of ports on the device (two ports on one side and two ports on the opposite side). Unfortunately, using $90^{0}$ probes decreases the calibration accuracy due to the poor planarity adjustment of probe tips that have a $90^{0}$ bend.

Measurements are made by connecting the network analyzer to 2 of the 4 ports of a component via a pair of calibrated microwave probes. The other two ports are contacted by probes that are terminated in 50~$\Omega$s. The network analyzer is then used to measure the S-parameters between the two reference planes of the two connected ports. This procedure is repeated for all pairs of probes that that can contact opposite sides of a device. In this manner the full 4$\times$4 S-matrix is determined. An important limitation of this technique arises because pairs of probes on the same side of a device can not be calibrated as the TRL standards themselves have probe pads on opposite sides. In addition, this procedure assumes that the unmeasured ports are in fact terminated in exactly 50~$\Omega$s.

\section{Results and Analysis}

Fig.~\ref{fig_DirCouAm}(b) shows the calibrated measurements of the directional coupler. The coupling is around $-19$~dB in a 2~GHz band around 6~GHz. The isolation is $> 20$~dB and the return loss $> 30$~dB in the same band. There is a weak resonant feature around 7.3~GHz, whose origin is unknown. We believe that this is a real resonance in the physical geometry of the coupler, not just a flaw in the calibration. We infer the quality of the calibration from the oscillations in the return loss. The period of these oscillations (400 MHz) is too small to be caused by reflections on the chip, rather they almost certainly result from a small standing wave component in the measurement cables which is not perfectly calibrated away. The observed return loss of 30 dB is consistent with the specified 30 dB return loss of the microwave probes used to terminate the two unmeasured ports. Fig.~\ref{fig_QuaHybAm}(b) shows the calibrated measurements of the quadrature hybrid. The isolation $> 20$~dB in a 700~MHz band and the return loss $> 20$~dB in a 550~MHz band around 6.5~GHz. The through transmission and coupling are $-3.1$~dB and $-3.3$~dB, respectively.

Measurements from both components show insertion loss $< 0.3$~dB. 
We believe that the $\approx0.2$~dB insertion loss that we do observe is not a consequence of power absorbed in the dielectric or conductors because the intrinsic loss of the cofabricated resonator yields a $Q > 10,000$. Losses might arise from coupling of the CPW modes to other modes, (including to free space modes), especially at the elbow structures. More striking is the fact that we also observe that the hybrid has its center frequency 500~MHz below the design value. This center frequency shift is likely caused by a poorly modeled reactance at the T-junctions in the CPW structures. This poorly modeled reactance may arise from dividing the entire hybrid structure into smaller parts for simulation.

\section{Conclusion}
We have designed, simulated, and measured two superconducting, passive, microwave components: a quadrature hybrid, and a 20~dB directional coupler. The performance of these components is within the design goals of a specific quantum information processing architecture known as continuous variables\cite{Braunstein:RevModPhys:2005}. However, we observe a discrepancy between the designed and measured center frequency of the hybrid. Nevertheless, both devices are appropriate for integration with JPAs whose center frequency can by widely tuned\cite{Manuel:NaturePhys:2008}. Furthermore, they may well be suited for use in other experiments that study the quantum optics of microwave photons\cite{Bozyigit:arxiv:2010}.


%



\section*{Acknowledgment}

The authors would like to thank Nathan Orloff and James Booth for discussions and very helpful advice.

\ifCLASSOPTIONcaptionsoff
  \newpage
\fi



\bibliographystyle{IEEEtran}
\bibliography{IEEEabrv,microwavebib}

\begin{thebibliography}{10}
\bibitem{nonendorsment}
The identification of specific commercial instruments does not imply
  endorsement by NIST nor does it imply that the instruments identified are the
  best available for a particular purpose.
\end{thebibliography}
\end{document}